# Security Document Classification with a Fine-Tuned Local Large Language Model: Benchmark Data and an Open-Source System

Ivan Dobrovolskyi – MS in AI & ML Engineering, Independent Researcher, Sunnyvale, CA, USA. ORCID: https://orcid.org/0009-0005-6938-0384

## Highlights

- Local model reached 95.0% accuracy in security document classification
- Open-source system screens sensitive files without cloud upload
- Training set includes 78,358 samples from 13 openly licensed sources and GPT-4 synthetic data
- Held-out evaluation supported performance beyond the main benchmark
- False positives stayed low on safe documents in the primary benchmark

## Abstract

Organizations that scan documents for sensitive information face a practical problem. Cloud services require data to be sent to external infrastructure, while rule-based tools often miss threats that depend on context. This study presents TorchSight, an open-source local system for security document classification built around a fine-tuned Qwen 3.5 27B model. The model was trained on 78,358 samples from 13 permissively licensed sources and GPT-4 synthetic data covering seven security categories and 51 subcategories. In the main evaluation on 1,000 documents, the model reached 95.0% category-level accuracy (95% confidence interval: 93.5–96.2). The tested commercial models scored 75.4-79.9% under the same prompting protocol. On a separate external set of 500 held-out samples, the model reached 93.8% accuracy, which suggests that performance extends beyond the main benchmark, although the margin depends on dataset composition and difficult boundary cases. The results show that a fine-tuned local model can support accurate security document classification while keeping document processing under local control.

## 1. Introduction

Data Loss Prevention (DLP) remains a practical challenge in security work. Organizations in healthcare, finance, government, and other regulated sectors rely on DLP tools to reduce the risk of data leaks. Most current approaches rely on regular expressions, keyword lists, and related rule-based techniques (Arp et al., 2021; Thomas et al., 2015). These methods work well for fixed patterns. They are less reliable when classification depends on context. Rule-based systems often fail to separate real sensitive content from instructional, placeholder, or illustrative material. The same pattern may appear in harmless documentation and in an actual exposure. Similar errors can also appear in prompted general-purpose models when safe technical content is interpreted as evidence of credential exposure. Sensitive information is not always marked by stable textual patterns. Medical, military, and financial documents may contain restricted content that requires contextual interpretation. Pattern matching alone is often insufficient for reliable classification. The same limitation appears in newer attack types. Prompt injection, supply chain poisoning, and server-side template injection do not follow a single fixed syntactic structure (Greshake et al., 2023; Ohm et al., 2020). Their detection depends on context and intended function rather than on surface form alone. In practice, organizations often combine multiple specialized tools, such as TruffleHog for secrets (Truffle Security, 2022), Semgrep for code vulnerabilities (R2C, 2020), and custom regular expressions for PII. This creates extra maintenance work and leaves classification criteria inconsistent across tools.

Cloud-based large language models such as GPT-5 (OpenAI, 2025), Claude (Anthropic, 2025), and Gemini (Gemini Team, 2023) could address all four limitations through their broad language understanding. Recent studies similarly note that large off-the-shelf language models perform strongly on general language and software-security tasks, yet their zero-shot performance on specialized cybersecurity tasks remains uneven (Jelodar et al., 2026; Huang, 2025). However, sending documents to a cloud API for security classification introduces the same data exposure risk that DLP is intended to prevent. This is a serious limitation for classified government systems, air-gapped military networks, healthcare environments, and organizations subject to data residency requirements. Even when providers offer contractual guarantees, documents still pass through external infrastructure during processing. This concern is consistent with privacy-oriented research showing that external processing of sensitive data in cloud environments introduces both compliance and trust risks (Baseri et al., 2024). These limitations point to a practical gap. Organizations need classification models that understand security-specific content, but they also need to keep sensitive documents under local control. This requirement is particularly important for enterprises, government-related users, healthcare providers, and security teams working under strict regulatory or data-residency constraints. For such users, accuracy alone is not enough. The classification process must also preserve privacy and prevent document exposure outside the local environment.

This study introduces TorchSight, an open-source system for local security document classification. The system uses domain-specific fine-tuning of a locally deployable large language model and is designed to operate without sending documents to cloud services.

The main contributions of the study are as follows:

1. We provide empirical evidence that domain-specific fine-tuning improves fixed-taxonomy security document classification. Under the evaluated protocol, the fine-tuned local model outperformed the tested prompted commercial models by 15–20 percentage points.

2. We propose a unified security taxonomy with 51 subcategories covering PII, credentials, financial records, medical data, confidential content, malicious payloads, and safe documents.

3. We present a 78,358-sample instruction-tuning dataset built from 13 public-domain and permissively licensed sources. The dataset was prepared through a reproducible pipeline and includes hard-negative boundary cases designed to test safe/unsafe discrimination.

4. We introduce TorchSight Beam, a Qwen 3.5 27B model fine-tuned with LoRA ($r = 128$, $\alpha = 256$). In the reported evaluation, Beam q4_K_M achieved 95.0% category-level accuracy while running on consumer hardware with 32 GB Apple Silicon memory.

5. We provide a complete open-source implementation, including a Rust CLI and a Tauri desktop frontend. The system runs fully locally and produces structured outputs in JSON, SARIF 2.1.0, HTML, and PDF formats.

## 2. Related Work

### 2.1. Rule-based secret and vulnerability detection

Existing secret-detection tools, including TruffleHog (Truffle Security, 2022), detect-secrets (Yelp, 2018), and GitLeaks (Zricethezav, 2019), mainly rely on regular expressions and entropy-based checks. This approach is effective when sensitive data follows a stable format, such as known API-key prefixes or fixed-length encoded tokens. However, it becomes less reliable in natural-language documents, where the same pattern may appear in examples, documentation, or harmless test data. As a result, these tools can generate many false positives and fail to capture threats that depend on context rather than form. Semgrep (R2C, 2020) adds abstract syntax tree (AST)-aware pattern matching, which improves code analysis. Still, it remains focused on syntactic rules and is not designed for semantic classification of natural-language documents.

### 2.2. LLM-based security analysis

SecureBERT (Aghaei et al., 2022) fine-tunes BERT for cybersecurity NLP tasks including named entity recognition. It targets information extraction from security text rather than document-level classification, and its encoder-only architecture cannot produce structured JSON output. CyberBench (Liu et al., 2024) benchmarks general-purpose LLMs on cybersecurity knowledge tasks; their results confirm that general-purpose models possess security knowledge but struggle with structured output and domain taxonomies – consistent with our finding that the base Qwen 3.5 27B reaches 86.3% category-level accuracy but only 19.0% on the 51-subcategory schema. LLM4Vuln (Sun et al., 2024) applies LLMs to code-level vulnerability detection, and PrivacyLens (Shao et al., 2024) demonstrates LLMs for privacy policy analysis. None provide a unified, on-premise model for cross-domain security document classification.

### 2.3. Data loss prevention systems

Commercial DLP solutions – Symantec DLP (Broadcom, n.d.), Microsoft Purview (Microsoft, 2026), Nightfall AI (Nightfall, n.d.) – combine regex pattern matching, machine learning classifiers, and cloud-based analysis. They require either cloud connectivity (creating the data exposure paradox where sensitive documents must be uploaded to a third-party service for classification) or expensive on-premise infrastructure with proprietary licensing. Nightfall AI specifically markets LLM-based DLP but operates as a cloud service, requiring document upload to their API – precisely the data egress that security-conscious organizations seek to avoid.
Open-source alternatives such as OpenDLP (Slaviero, 2012) and MyDLP (Comodo, 2014) rely exclusively on regex pattern matching and have not been actively maintained since 2015 and 2016 respectively. To the best of our knowledge, no actively maintained open-source DLP system currently offers comparable LLM-based cross-domain security document classification. TorchSight fills this gap: a single fine-tuned LLM replaces the regex + ML + cloud stack, runs locally on commodity hardware, and covers a broader category space than any single existing tool, commercial or open-source.

### 2.4. LoRA and parameter-efficient fine-tuning

Low-Rank Adaptation (Hu et al., 2021) enables efficient fine-tuning by injecting trainable low-rank matrices into transformer layers while keeping pre-trained weights frozen. QLoRA (Dettmers et al., 2023) further reduces memory through quantized base weights. We use standard LoRA with bf16 precision, as our training infrastructure provides sufficient memory for the full 27B model.

## 3. Security Taxonomy

We define a hierarchical taxonomy of 7 top-level categories and 51 subcategories designed to cover the full spectrum of document security concerns. The taxonomy draws on NIST SP 800-53 (NIST, 2020a), MITRE ATT&CK (MITRE Corporation, 2023), OWASP Top 10 (OWASP, 2021), CWE (MITRE Corporation, n.d.), and compliance frameworks including GDPR (European Parliament, 2016), HIPAA (U.S. Congress, 1996), PCI-DSS (PCI SSC, 2022), and ITAR (U.S. Department of State, 2024).

To operationalize document classification in a consistent and reproducible manner, the proposed security taxonomy is summarized in Table 1.

**Table 1.** Security taxonomy: 7 categories and 51 subcategories.

| Category | Sub. | Subcategories | Example Content |
|---|---|---|---|
| **Malicious** | 14 | injection, exploit, shell, phishing, malware, prompt_injection, supply_chain, ssti, ssrf, ... | SQL injection, reverse shells |
| **Confidential** | 9 | classified, internal, military, military_comms, intelligence, weapons_systems, ... | TOP SECRET markings, OPORDs |
| **Credentials** | 8 | password, api_key, token, private_key, connection_string, cloud_config, cicd, container | Hardcoded AWS keys, DB URLs |
| **PII** | 6 | identity, contact, government_id, biometric, behavioral, metadata | Names+SSN, fingerprints |
| **Safe** | 6 | documentation, code, config, media, email, business | Tutorials, open-source code |
| **Financial** | 4 | credit_card, bank_account, transaction, tax | Credit card numbers, W-2 forms |
| **Medical** | 4 | diagnosis, prescription, lab_result, insurance | Clinical notes, prescriptions |

**Source:** created by the author based on the proposed security taxonomy developed in the present study.

The taxonomy includes LLM-era attack vectors (prompt injection, supply chain poisoning) alongside traditional categories. Each finding carries a severity level (critical, high, medium, low, info) and maps to exactly one subcategory for unambiguous evaluation.

## 4. Dataset

### 4.1. Data sources

We curate training data from 13 sources, all verified as public domain (U.S. Government works under 17 U.S.C. § 105) or permissively licensed (Apache 2.0, MIT, CC-BY-4.0). The dataset excludes both copyleft-licensed sources (GPL/LGPL) and ShareAlike-licensed materials (CC BY-SA). As a result, it remains suitable for commercial training use without introducing additional licensing obligations. Table 2 presents the composition of the training corpus, including the original and rebalanced sample counts, licensing status, and category coverage for each source.

**Table 2.** Training data sources. Raw = before rebalancing; Balanced = final training set after per-subcategory capping at 5,000.

| Source | Raw | Balanced | License | Categories |
|---|---|---|---|---|
| **NVD CVE Database** | 50,000 | 8,475 | Public Domain | malicious.exploit |
| **Synthetic augmentation** | 33,100 | 39,754 | Generated | All categories |
| **Hard negatives** | 6,400 | 6,134 | Generated | Boundary cases |
| **AI4Privacy** | 5,000 | 4,851 | Apache 2.0 | pii.* |
| **Fenrir v2.0** | 5,000 | 4,573 | Apache 2.0 | malicious.* |
| **SEC EDGAR** | 3,000 | 3,000 | Public Domain | financial.* |
| **SecLists** | 3,229 | 1,708 | MIT | malicious.injection |
| **Phishing Dataset** | 3,000 | 2,796 | Apache 2.0 | malicious.phishing |
| **NIST Training** | 3,000 | 2,761 | Public Domain | safe.documentation |
| **Enron Email Corpus** | 2,000 | 1,902 | Public Domain | pii.*, credentials.* |
| **MITRE ATT&CK v14** | 1,620 | 871 | Royalty-free | malicious.malware |
| **Loghub** | 1,280 | 1,280 | Free for research | safe.config |
| **Other (3 sources)** | 327 | 253 | Various | Multiple |

| | | | permissive |
|---|---|---|---|
| **Total** | 116,956 | 78,358 | |

**Source:** created by the author based on the multi-source training corpus compiled and rebalanced in the present study.

Table 2 shows that the final dataset was built from multiple source types rather than from a single dominant corpus. During rebalancing, the largest sources were reduced in weight, while smaller categories were preserved to maintain coverage across the full taxonomy. This makes the corpus more suitable for domain-specific fine-tuning and also strengthens its practical usability, since the data comes from public-domain or permissively licensed materials.

### 4.2. Synthetic augmentation and hard negatives

To cover all 51 subcategories at a usable level, we generated 33,100 synthetic samples with GPT-4. Each sample was checked against the taxonomy schema and compared with the real-data subset to remove duplicates. We also created 6,400 hard-negative samples to test cases where safe and unsafe content are difficult to separate. This group included 3,000 dangerous documents that looked safe, 2,500 safe documents that looked dangerous, and 900 boundary cases, such as multi-category files and partially redacted documents. These examples were added because many security errors arise in such borderline cases. They helped train the model to look beyond surface patterns and consider the security meaning of the text in context.

### 4.3. AI-assisted data generation disclosure

Following Elsevier's guidance on AI-assisted technologies, we report that GPT-4 (OpenAI, 2023) was used only for generating synthetic and hard-negative training samples. The first generation stage produced 33,100 synthetic samples and 6,400 hard-negative samples. During dataset rebalancing, additional GPT-4-generated samples were added to subcategories with insufficient representation. In the final 78,358-sample training set, synthetic augmentation accounts for 39,754 samples (50.7%), while hard-negative samples account for 6,134 samples (7.8%). All generated samples were checked against the taxonomy schema and deduplicated against the real-data subset.

### 4.4. Rebalancing and instruction format

The initial dataset contained 116,956 samples and was unevenly distributed across categories. The largest imbalance came from the NVD (NIST, 2020b) vulnerability descriptions, which accounted for 50,000 samples. To reduce this skew, we limited each subcategory to a maximum of 5,000 samples and added synthetic examples to categories that remained underrepresented. The resulting balanced training set contained 78,358 samples, including 74,441 training samples and 3,917 validation samples under a 95/5 split. Table 2 reports the source counts before and after rebalancing.

All samples were formatted in the Alpaca instruction style with a fixed system prompt specifying the classification task, the output schema (a JSON array of findings), and the category taxonomy. During SFT conversion, the user instruction was randomly selected from seven equivalent phrasings of the same request in order to make the model less dependent on a single prompt wording.

## 5. Model

### 5.1. Base model selection

We used Qwen 3.5 27B (Qwen Team, 2026) as the base model. Its size offered a practical compromise between model capability and local deployability. In our setup, the q4_K_M quantization (approximately 17 GB) ran at interactive speed on a single Apple Silicon Mac with 32 GB of unified memory, whereas q8_0 (approximately 28 GB) was better suited to discrete GPUs with at least 48 GB of VRAM.

### 5.2. LoRA training configuration

The configuration of the LoRA fine-tuning procedure, including model adaptation targets, optimization settings, and hardware environment, is presented in Table 3.

**Table 3.** LoRA training hyperparameters.

| Parameter | Value |
|---|---|
| **LoRA rank (r)** | 128 |
| **LoRA alpha (α)** | 256 |
| **Target modules** | q/k/v/o_proj, gate/up/down_proj |
| **Dropout** | 0.05 |
| **Learning rate** | $2 \times 10^{-5}$ |
| **Scheduler** | Cosine decay, 10% warmup |
| **Effective batch size** | 16 (4 × 4 gradient accumulation) |
| **Epochs** | 5 |
| **Precision** | bf16 |
| **Max sequence length** | 4,096 tokens |
| **Hardware** | 8× NVIDIA A100 80GB SXM4 |

**Source:** created by the author based on the LoRA training configuration used in the present study.

As shown in Table 3, the training setup combines parameter-efficient adaptation with a relatively high-capacity optimization regime, enabling domain specialization of the 27B base model while maintaining computational feasibility.

Training used the Hugging Face TRL library (v0.11.4) with SFTTrainer, transformers v4.45.2, and PEFT v0.13.2. The trained adapter is merged with base weights and exported to GGUF format via llama.cpp (Gerganov, 2023) at three quantization levels. Distribution is through Ollama (Ollama, n.d.) for single-command installation. Inference uses temperature = 0 for deterministic classification.

## 6. **Materials and Methods**

### 6.1. **Research design and procedure**

This study adopted a controlled comparative design to evaluate whether domain-specific fine-tuning improves security document classification relative to prompted frontier models, a rule-based baseline, and the untuned base model. The research procedure comprised four stages: (1) development of a hierarchical security taxonomy, (2) construction and rebalancing of a multi-source training dataset, (3) LoRA fine-tuning of Qwen 3.5 27B, and (4) standardized evaluation on internal and external benchmarks under matched inference conditions. This design made it possible to isolate the contribution of domain adaptation while preserving comparability across models.

### 6.2. Evaluation independence and data separation

To reduce the risk of evaluation leakage, the supervised fine-tuning data, the primary benchmark, and the external validation set were treated as separate resources. The primary benchmark was generated after model training and checked against the training corpus to exclude exact duplicates, whereas the external validation set was assembled from held-out and explicitly excluded sources. This separation does not eliminate all possible distributional overlap, especially where synthetic data are involved, but it strengthens the interpretation of both the benchmark and generalization results.

### 6.3. Experimental setup

All models receive identical treatment: the same system prompt, input format (documents truncated to 6,000 characters), and JSON parser. Local models and Claude/Gemini use temperature = 0 for deterministic inference. GPT-5 is an exception, as it only supports temperature = 1. Because GPT-5 does not allow the same deterministic inference setting as the other evaluated systems, its results should be interpreted as approximate under the present protocol rather than as a perfectly matched estimate. To preserve procedural comparability, all evaluated models were run with the same task prompt, input format, and JSON parsing pipeline. This choice should not be interpreted as an optimized comparison for every commercial model. Rather, it evaluates how well prompted general-purpose models follow a shared taxonomy-constrained protocol without model-specific adaptation. Accordingly, the results compare a domain-adapted model against prompted baseline systems under a common evaluation interface, not against vendor-optimized prompting strategies. GPT-5 also requires the max_completion_tokens parameter instead of the standard max_tokens, reflecting a distinct API contract relative to the other evaluated models. When multiple findings are present, the most severe non-safe finding determines the primary category. To ensure a fair comparison, all model evaluations were performed without the regex-based detection layer. This allows isolating the contribution of the LLM component. The impact of hybrid approaches (LLM + regex) is evaluated separately. All experiments were conducted using the same evaluation pipeline and input data, ensuring full comparability and reproducibility of results.

### 6.4. Benchmark dataset and sample

We evaluate on two complementary benchmarks. **TorchSight-Eval-500-External** draws 500 documents from public datasets — 100 NVD CVEs, 80 NIST cybersecurity publications, 100 MTSamples clinical transcriptions, 80 AI4Privacy PII records, 80 Enron emails, and 60 phishing emails — covering four taxonomy categories (PII, medical, malicious, safe) where high-quality public data is available. **TorchSight-Eval-1000** is a controlled synthetic benchmark of 1,000 documents stratified across all seven categories: credentials (n = 150), PII (n = 150), malicious (n = 150), safe (n = 250), financial (n = 100), medical (n = 100), and confidential (n = 100). Safe documents are deliberately overrepresented to stress-test false-positive behavior, which is operationally critical for DLP.

The two benchmarks are complementary, not redundant. The external set validates generalization to real-world distributions on the four categories where public data exists. The synthetic set provides (a) coverage of three categories — credentials, financial, and confidential — for which no public benchmark exists, by structural data-availability constraints (leaked credential corpora cannot be lawfully redistributed; sensitive operational financial documents are regulated and proprietary; classified content is by definition unpublished), and (b) stratified borderline-case coverage including hard-negative-style boundary documents designed to test safe/unsafe discrimination, which public datasets do not expose at controllable quantity. Together the two benchmarks provide both real-data validation and stratified per-class measurement across the full taxonomy.

### 6.5. Evaluation metrics and instruments

Because the task is a multiclass document classification problem, each document was assigned a single primary category label based on the most severe detected non-safe finding. Overall model performance was assessed using category-level accuracy, calculated as the proportion of correctly classified documents among all evaluated samples. In addition, precision, recall, and F1-score were computed for each category using a one-vs-rest scheme and then macro-averaged across categories to provide a balanced estimate of class-wise performance.

Accuracy evaluates the overall classification performance by measuring the proportion of correctly classified instances among all samples:

$$\text{Accuracy} = \frac{\text{Number of correctly classified samples}}{\text{Total number of samples}} \quad (1)$$

Precision measures the proportion of correctly predicted positive instances among all instances predicted as positive:

$$\text{Precision} = \frac{TP}{TP+FP} \quad (2)$$

Recall evaluates the model's ability to correctly identify positive instances among all actual positive cases:

$$\text{Recall} = \frac{TP}{TP+FN} \quad (3)$$

F1-score represents the harmonic mean of precision and recall, providing a balanced measure of model performance:

$$F1 = 2 \times \frac{\text{Precision} \times \text{Recall}}{\text{Precision} + \text{Recall}} \quad (4)$$

where:
TP – true positives;
FP – false positives;
FN – false negatives.

Precision, recall, and F1-score were computed for each category using a one-vs-rest scheme, while accuracy was calculated at the overall category level across all evaluated samples. Each document was assigned a single primary label based on the most severe detected non-safe category. All metrics were computed using a unified evaluation script to ensure consistency across model comparisons.

### 6.6. System Architecture

The overall architecture of the proposed TorchSight system is presented in Figure 1.

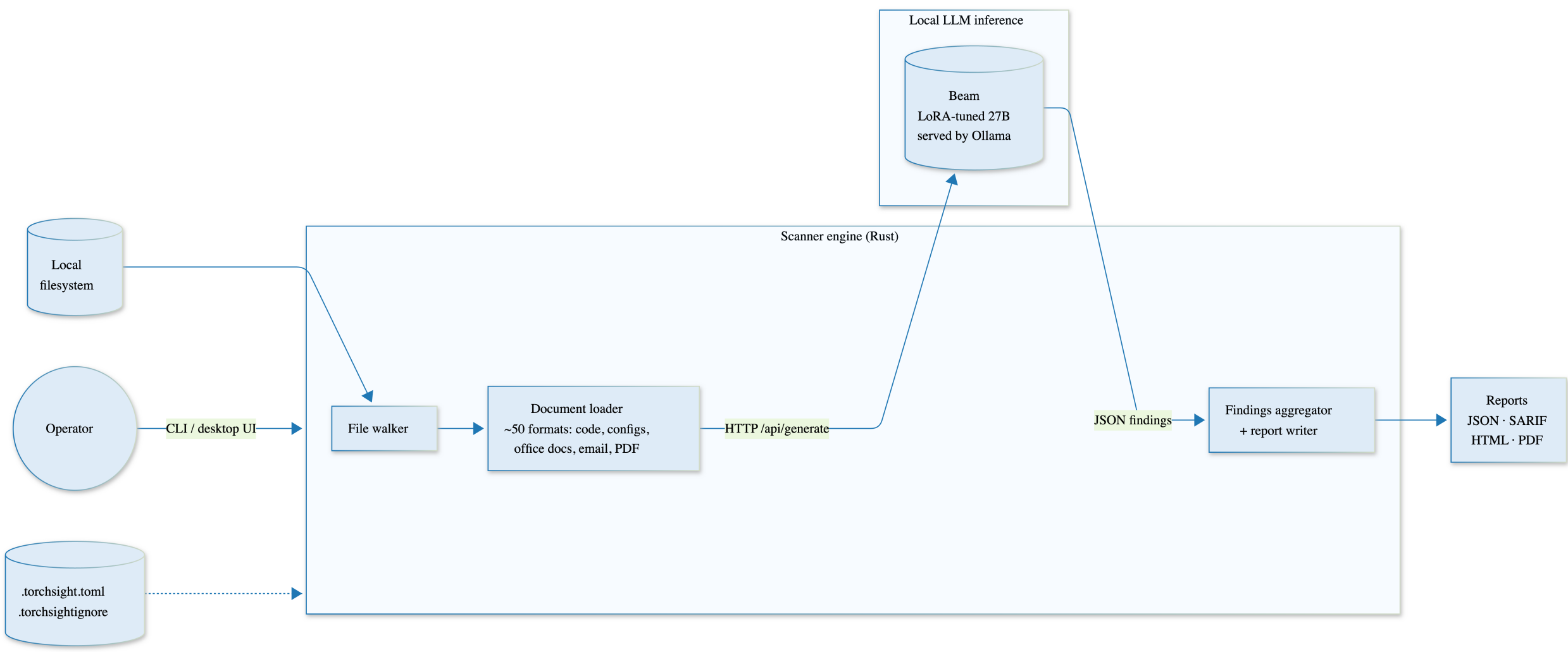

**Figure 1.** TorchSight system architecture. The Rust scanner engine reads documents from the filesystem and queries the local Beam model through the local Ollama HTTP API; outputs are written as structured reports in JSON, SARIF 2.1.0, HTML, and PDF formats.

**Source:** created by the author based on the TorchSight system design and implementation in the present study.

TorchSight is implemented in Rust. The scanner recursively walks the target directory, respecting .torchsightignore exclusion patterns. The loader supports approximately 50 document and code formats, including plain text, configuration files, office documents, emails, and PDFs. Text files are read directly and truncated to 6,000 characters; PDFs are converted to text via pdftotext.

Communication with Beam uses Ollama's local HTTP API (/api/generate) in Alpaca instruction format. A critical implementation detail: Beam requires a raw prompt template rather than a chat template. Using the chat endpoint caused autocompletion instead of classification, dropping accuracy to near zero. Forty-eight compiled regex patterns provide a deterministic fallback layer. The system outputs JSON, SARIF 2.1.0, HTML, and PDF reports. CI/CD integration includes --fail-on threshold exits, --stdin pipe support, --diff scanning. A Tauri v2 (Tauri, 2024) desktop application provides a GUI for security analysts and compliance officers.

TorchSight was designed to preserve privacy at the architectural level rather than relying only on policy commitments. During scanning, the Rust binary performs no external network communication. Its only inference call is to the local Ollama endpoint (/api/generate) on localhost; no document content is transmitted to cloud services. The scanner includes no telemetry or update-checking component in the scanning path. All reports are written to local storage.

This design makes the system suitable for fully local deployment, including air-gapped environments, healthcare settings, and infrastructures subject to data-residency requirements. In contrast to cloud-based DLP services, where privacy depends on vendor-controlled processing, TorchSight keeps document analysis within the local environment by design. These implementation choices are also important for interpreting the system's practical advantages relative to existing approaches.

### 6.7. Ethical considerations

The dataset includes real vulnerability descriptions and attack payloads from public sources. To limit misuse risk, the system was designed for classification only, using structured JSON output and a constrained instruction-tuning setup. It is released under Apache 2.0 with responsible-use documentation, and inference is performed at temperature 0 to keep generation within the classification task.

## 7. Results

We report results on two complementary benchmarks introduced in Section 6.4. The synthetic TorchSight-Eval-1000 (7.1–7.4) provides stratified per-class measurement across all seven categories, including the three (credentials, financial, confidential) for which no public benchmark exists, and includes boundary cases designed to test safe/unsafe discrimination. The held-out TorchSight-Eval-500-External (7.5) provides real-world distribution validation on the four categories where public data is available (PII, medical, malicious, safe). Read together, the two benchmarks deliver controlled coverage and external-data validation; readers primarily interested in real-world generalization may skip directly to Section 7.5. The 7.6 reviewer-validation subsection then quantifies benchmark label fidelity through independent blind re-annotation.

### 7.1. Overall benchmark performance

The benchmark results demonstrate a clear advantage of the domain-specific fine-tuned Beam models over both frontier commercial APIs and non-fine-tuned baselines. Overall classification performance across all evaluated systems is summarized in **Table 4**.

**Table 4.** Overall classification accuracy (1,000 text samples, temperature = 0 except GPT-5). 95% Wilson score CIs.

| Model | Type | Cat. Acc. [95% CI] | Subcat. | Time |
|---|---|---|---|---|
| **Beam q4_K_M** | Local (LoRA) | 95.0% [93.5, 96.2] | 48.2% | 4.4 s |
| **Beam f16** | Local (LoRA) | 93.2% [91.5, 94.6] | 51.1% | 4.6 s |
| **Beam q8_0** | Local (LoRA) | 93.0% [91.2, 94.4] | 51.4% | 3.2 s |
| **Claude Sonnet 4** | Commercial API | 79.9% [77.3, 82.3] | 23.0% | ~6 s |
| **Claude Opus 4** | Commercial API | 79.9% [77.3, 82.3] | 22.5% | ~13 s |
| **GPT-5** | Commercial API | 76.9% [74.2, 79.4] | 11.6% | ~24 s |
| **Gemini 2.5 Pro** | Commercial API | 75.4% [72.6, 78.0] | 21.0% | ~14 s |
| **Regex-only (48 pat.)** | Rule-based | 52.7% [49.6, 55.8] | – | 0.2 ms |
| **Qwen 3.5 27B base** | Local (no LoRA) | 86.3% [84.0, 88.3] | 19.0% | ~40 s |

**Source:** created by the author based on the results of the primary TorchSight-Eval-1000 benchmark.

As shown in Table 4, all three Beam quantizations (93.0–95.0%) outperform every commercial frontier model (75.4–79.9%) by 13–20 percentage points. Confidence intervals do not overlap between any Beam variant and any commercial model. GPT-5 achieves 76.9% on the primary benchmark, placing it between the Claude models and Gemini. By comparison, fine-tuning yields an 8.7 pp category-level gain over the base Qwen 3.5 27B model (95.0% vs 86.3%) and a 29.2 pp gain on subcategory adherence (48.2% vs 19.0%). At the same time, this margin should be interpreted in light of the evaluation protocol: Beam was optimized for the present taxonomy and output format, whereas the commercial models were evaluated as prompted general-purpose baselines without model-specific tuning. The reported advantage therefore supports a benchmark-specific claim about the value of domain adaptation, rather than a universal claim of superiority across all cybersecurity tasks. However, aggregate benchmark accuracy alone does not show whether Beam's advantage is evenly distributed across security classes or driven by a subset of categories. For this reason, the next subsection disaggregates the results by category and examines false-positive behavior on benign documents.

### 7.2. Per-category performance and false positives

To examine whether this overall advantage is consistent across specific security classes, the category-level accuracy breakdown is presented in Table 5.

**Table 5.** Per-category accuracy breakdown (%).

| Category (n) | Beam q4_K_M | GPT-5 | Sonnet 4 | Opus 4 | Gemini 2.5 |
|---|---|---|---|---|---|
| **Credentials (150)** | 96.0 | 99.3 | 100.0 | 100.0 | 100.0 |
| **PII (150)** | 100.0 | 88.7 | 90.0 | 87.3 | 89.3 |
| **Financial (100)** | 100.0 | 63.0 | 61.0 | 63.0 | 63.0 |
| **Medical (100)** | 68.0 | 48.0 | 40.0 | 55.0 | 80.0 |
| **Confidential (100)** | 100.0 | 100.0 | 99.0 | 61.0 | 85.0 |
| **Malicious (150)** | 95.3 | 98.7 | 98.0 | 96.7 | 100.0 |
| **Safe (250)** | 98.0 | 51.2 | 66.8 | 77.6 | 36.8 |

**Source:** created by the author based on the category-level results of the primary TorchSight-Eval-1000 benchmark.

As shown in Table 5, Beam q4_K_M performs particularly strongly on PII (100.0% vs. 87.3–90.0%), financial data (100.0% vs. 61.0–63.0%), and safe documents (98.0% vs. 36.8–77.6%). The most operationally important difference concerns the safe category, where false positives directly affect the usability of a DLP system. This contrast is illustrated more explicitly in Figure 2.

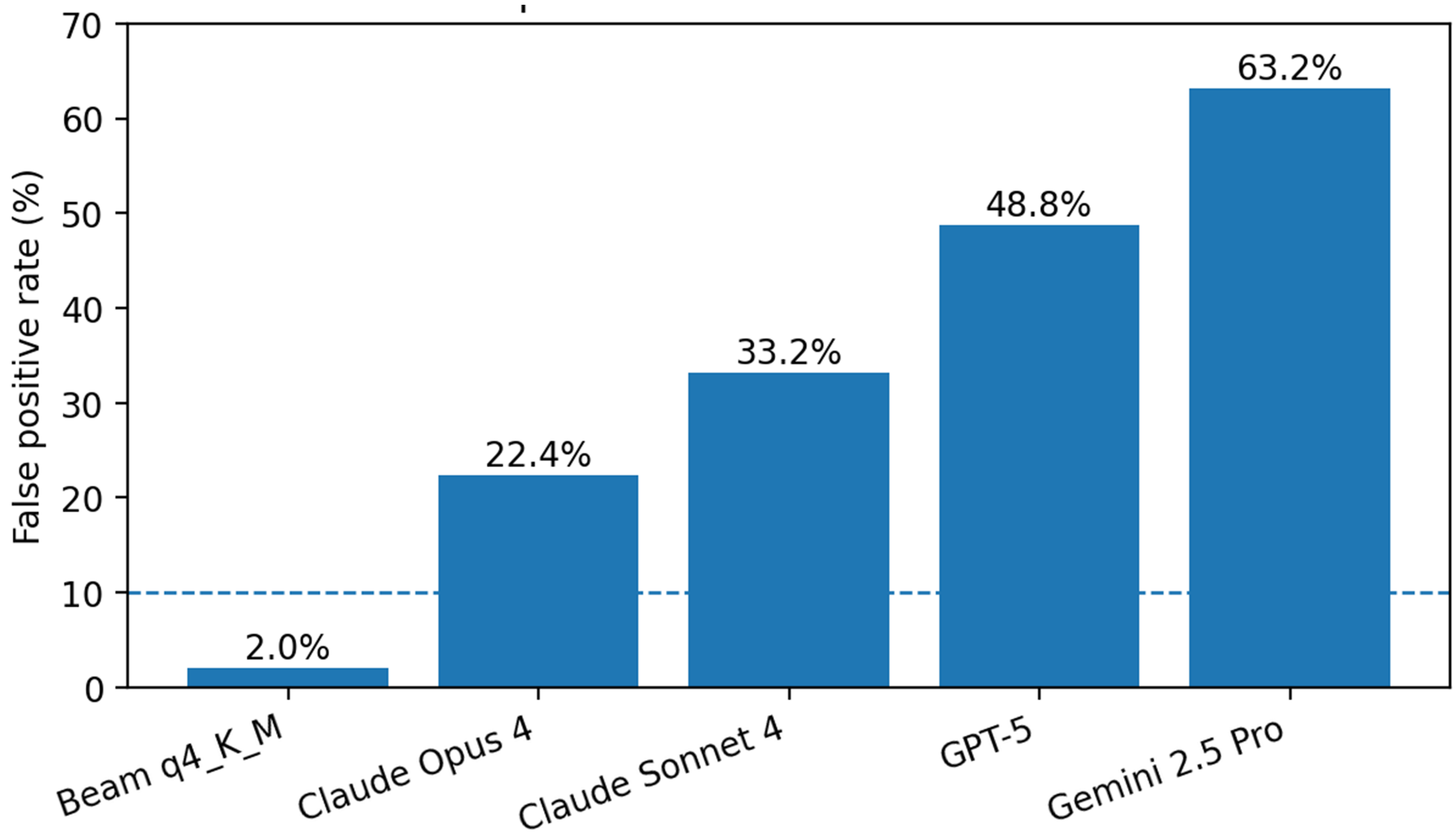

**Figure 2.** False positive rates on safe documents across the evaluated models in the primary benchmark.

**Source:** created by the author based on false positive rate calculations for safe documents in the primary benchmark.

As shown in Figure 2, only Beam remains below the 10% usability threshold, with a false positive rate of 2.0%, whereas Claude Opus 4 reaches 22.4%, Claude Sonnet 4 33.2%, GPT-5 48.8%, and Gemini 2.5 Pro 63.2%. In practical deployment settings, such false positive rates would make most commercial models difficult to use for routine document screening despite their otherwise competitive performance in selected categories. At the same time, all four commercial models lead Beam slightly on malicious detection (96.7–100.0% vs. 95.3%) and on credentials (99.3–100.0% vs. 96.0%), which may reflect broader prior exposure to security-oriented patterns. Medical classification remains the most challenging category overall, with Beam achieving 68.0%, Gemini 80.0%, and GPT-5 48.0%, suggesting semantic overlap between medical, PII, and confidential content.

To provide a more detailed assessment of Beam's class-wise behavior beyond accuracy alone, its precision, recall, and F1-score are reported in Table 6.

**Table 6.** Precision, recall, and F1 for Beam q4_K_M.

| Category | Precision | Recall | F1 |
|---|---|---|---|
| **Credentials** | 100.0% | 96.0% | 98.0% |
| **PII** | 87.2% | 100.0% | 93.2% |
| **Financial** | 100.0% | 100.0% | 100.0% |
| **Medical** | 100.0% | 68.0% | 81.0% |
| **Confidential** | 90.9% | 100.0% | 95.2% |
| **Malicious** | 92.9% | 95.3% | 94.1% |
| **Safe** | 97.2% | 98.0% | 97.6% |

| Macro average | 95.5% | 93.9% | 94.1% |
|---|---|---|---|

**Source:** created by the author based on precision, recall, and F1-score calculations from the primary benchmark evaluation.

Table 6 provides a more detailed view of Beam q4_K_M's performance by category. The model shows strong class-wise results, including 100.0% precision for credentials, financial documents, and medical documents. At the macro level, it reaches 95.5% precision, 93.9% recall, and 94.1% F1. The main weakness remains medical content: recall for this category is 68.0%, which means that part of the medical documents was assigned to other sensitive classes. These results suggest that Beam q4_K_M's advantage is not only a matter of overall benchmark accuracy. The model also maintains high discrimination across most security categories and produces the lowest false-positive rate on safe documents, an important requirement for practical DLP use. Figure 3 shows the same pattern in the confusion matrix. Six of the seven categories are classified perfectly or almost perfectly. The only clear concentration of errors appears in the Medical row, where 22 of 100 medical documents are classified as PII and another 10 as Confidential.

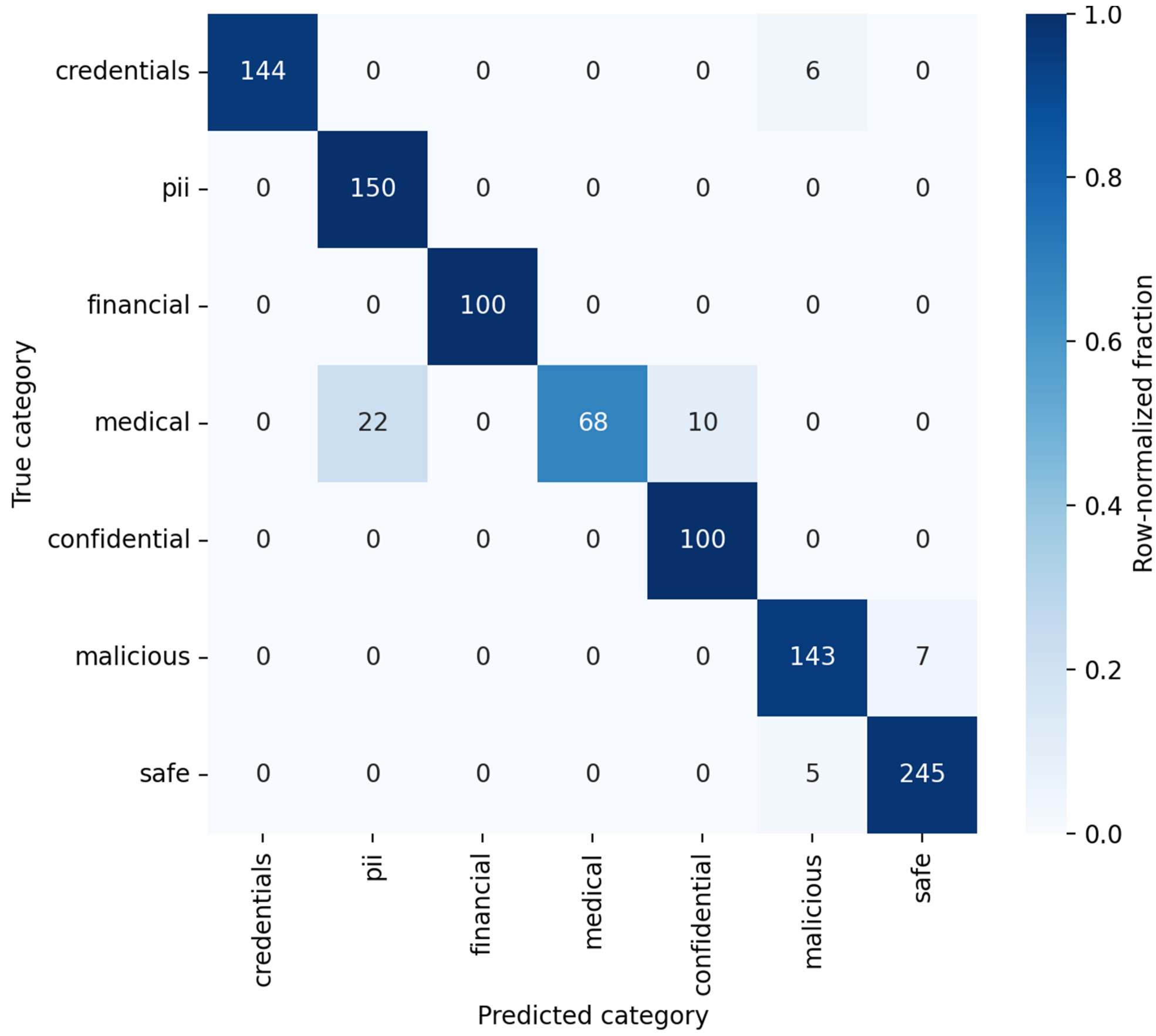

**Figure 3.** Confusion matrix for Beam q4_K_M on the 1,000-document primary benchmark. Cells are row-normalized; absolute counts annotated.

**Source:** created by the author based on the per-document predictions of Beam q4_K_M on the primary TorchSight-Eval-1000 benchmark. Rows represent true categories, and columns represent predicted categories. Cells are row-normalized, while the overlaid numbers show absolute counts. PII = personally identifiable information.

After examining Beam q4_K_M's performance by category, we next evaluate why this advantage appears. The key question is whether the improvement comes mainly from domain-specific fine-tuning or from other implementation choices, such as quantization. To address this, we compare Beam with the untuned Qwen 3.5 27B base model and then examine the results across the different quantization variants.

### 7.3. Fine-tuning, quantization, and subcategory behavior

The comparison between Beam q4_K_M and the untuned Qwen 3.5 27B model serves as an ablation test. Both models use the same underlying architecture, and the base model was evaluated with the same system prompt. The main difference appears at the subcategory level. Although the base model reaches 86.3% category-level accuracy, it achieves only 19.0% accuracy on the 51-subcategory schema. In many cases, it produces labels that are not part of the predefined taxonomy. This suggests that fine-tuning is especially important for enforcing the required taxonomy and output structure. Prompting alone was not sufficient to achieve the same level of schema adherence. An additional finding is that q4_K_M quantization outperforms both q8_0 and f16 at the category level (95.0% vs. 93.0% and 93.2%), whereas the higher-precision variants achieve slightly better subcategory accuracy (51.1–51.4% vs. 48.2%). The category-level gap (8.7 pp) and the subcategory gap (29.2 pp) together indicate that LoRA fine-tuning's primary contribution is not category discrimination — which a properly-prompted 27B base already handles well — but adherence to the prescribed 51-subcategory taxonomy. GPT-5 achieves the lowest subcategory accuracy at 11.6%, inventing its own subcategory names (e.g., credentials.aws_access_key_id, confidential.mnpi, medical.deidentified) rather than adhering to the specified taxonomy. Even its strong reasoning capabilities are therefore insufficient to enforce consistent use of a 51-subcategory schema from prompting alone. By contrast, Beam variants achieve 48.2–51.4% subcategory accuracy, outperforming commercial models by approximately 2–4 times. Because real-world DLP workflows still rely heavily on deterministic filters, model-level comparisons should also be complemented by a direct contrast with rule-based detection. The next subsection therefore evaluates the best-case regex-only baseline against Beam on the same benchmark.

### 7.4. Regex-only baseline

To quantify the gap between rule-based and LLM-based classification, we evaluate TorchSight's 48 compiled regex patterns in isolation (no LLM). These patterns cover structured indicators: AWS key prefixes, credit card Luhn patterns, SSN formats, known injection signatures, and private key headers. This represents a strong regex-style baseline for document classification, although it should not be interpreted as a full optimized run of each individual tool.

**Table 7.** Regex-only vs. Beam q4_K_M accuracy on Eval-1000.

| Category | Regex-only | Beam q4_K_M | Gap |
|---|---|---|---|
| **Credentials** | 84.0% | 96.0% | +12 pp |
| **Safe** | 87.2% | 98.0% | +11 pp |
| **PII** | 52.0% | 100.0% | +48 pp |
| **Malicious** | 38.0% | 95.3% | +57 pp |
| **Financial** | 48.0% | 100.0% | +52 pp |
| **Confidential** | 0.0% | 100.0% | +100 pp |
| **Medical** | 0.0% | 68.0% | +68 pp |

| **Overall** | 52.7% | 95.0% | +42.3 pp |
|---|---|---|---|

**Source:** created by the author based on the comparative evaluation of the regex-only baseline and Beam q4_K_M on TorchSight-Eval-1000.

Regex achieves 0% on confidential and medical content – simple regular-expression matching cannot reliably determine whether a military OPORD or a patient diagnosis constitutes sensitive information. Even on categories with structured patterns (credentials, financial), regex lags by 12–52 percentage points because it cannot handle context-dependent content: it flags tutorial placeholder keys as real credentials and misses semantically expressed secrets. The 42.3 pp overall gap between regex (52.7%) and Beam (95.0%) demonstrates that LLM-based classification provides fundamentally different capabilities than pattern matching.

### 7.5. External validation

Per-model accuracy on TorchSight-Eval-500-External is reported in Table 8.

**Table 8.** External validation: all models on 500 held-out samples.

| Source | n | Beam q4_K_M | Claude S4 | Gemini 2.5 | GPT-5 | Qwen base |
|---|---|---|---|---|---|---|
| **NVD held-out** | 100 | 100.0% | 98.0% | 97.0% | 51.0% | 97.0% |
| **NIST held-out** | 80 | 100.0% | 92.5% | 91.2% | 88.8% | 95.0% |
| **MTSamples** | 100 | 82.0% | 100.0% | 100.0% | 100.0% | 100.0% |
| **AI4Privacy** | 80 | 100.0% | 71.2% | 63.7% | 65.0% | 67.5% |
| **Phishing** | 60 | 100.0% | 55.0% | 45.0% | 28.3% | 68.3% |
| **Enron** | 80 | 83.8% | 87.5% | 77.5% | 47.5% | 81.2% |
| **Overall** | 500 | 93.8% | 86.4% | 82.0% | 65.8% | 86.6% |

**Source:** created by the author based on the held-out external validation benchmark constructed for the present study.

Beam q4_K_M's source-level performance on the external validation benchmark is visualized in Figure 4. Beam q4_K_M achieves 93.8% on the external benchmark.

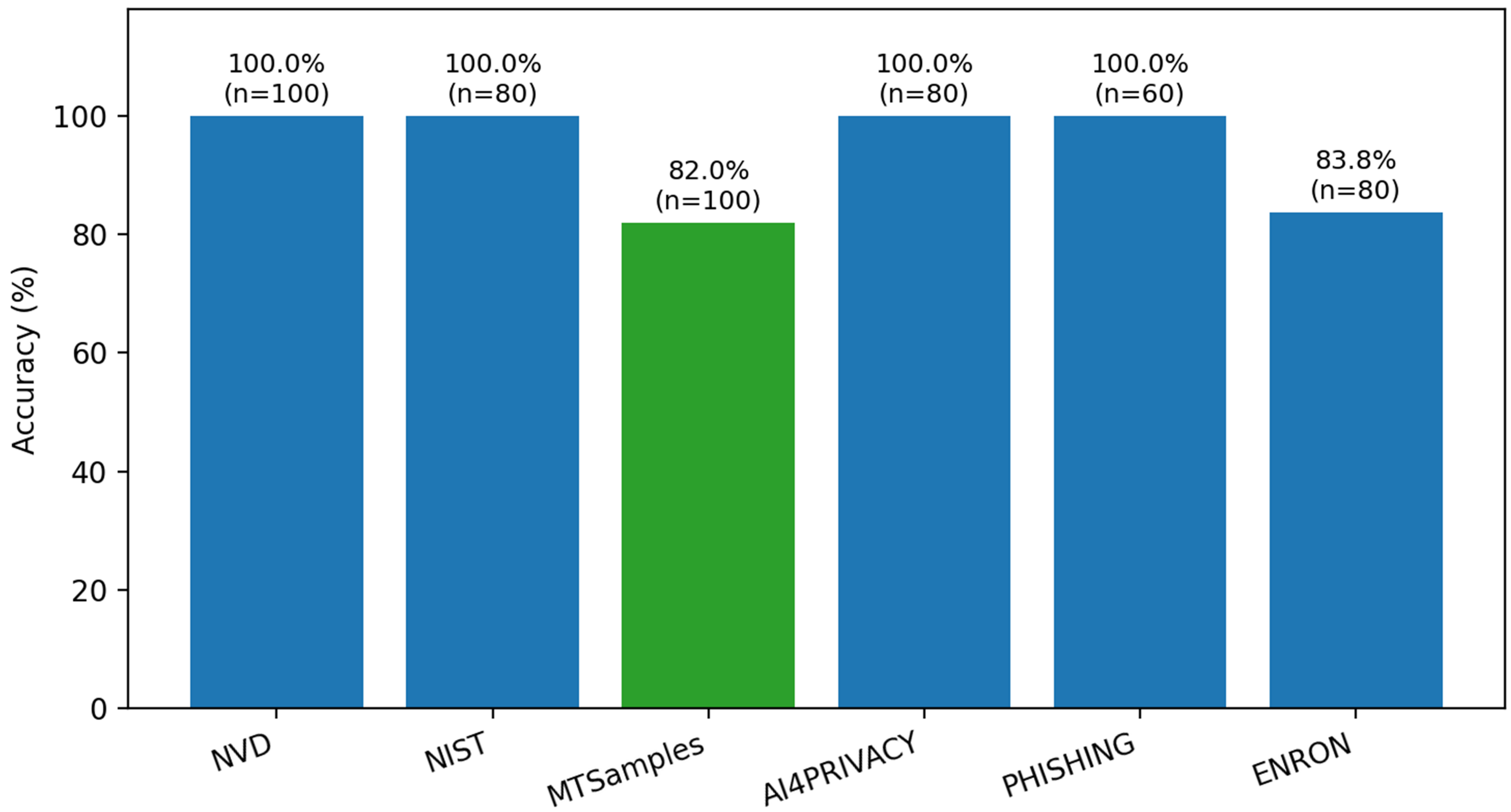

**Figure 4.** External validation: Beam q4_K_M accuracy by source on 500 held-out samples.

**Source:** created by the author based on source-level results from the held-out external validation benchmark.

Green bar highlights MTSamples (82.0%), a dataset explicitly excluded from training. Beam achieves 93.8% on the external benchmark, outperforming Claude Sonnet 4 (86.4%), Gemini 2.5 Pro (82.0%), and GPT-5 (65.8%). Claude, Gemini, GPT-5, and the Qwen base model all reach 100% on MTSamples – medical transcriptions that Beam never encountered during training – while Beam reaches 82%. The 18 misses route to the PII category rather than medical, with explanations that explicitly cite HIPAA, indicating the model recognizes the protected health information but assigns a different label under the present taxonomy. The largest model divergence appears on phishing emails (Beam 100%, Claude 55%, Gemini 45%, GPT-5 28.3%) and NVD vulnerability descriptions (Beam 100%, GPT-5 51%). Claude outperforms Beam on Enron corporate emails (87.5% vs. 83.8%), suggesting higher PII/confidential boundary ambiguity in this source.

The comparison between the primary and external benchmarks reveals a notable shift in model behavior. Claude Sonnet 4 and Gemini 2.5 Pro improve on the external dataset by +6.5 and +6.6 percentage points, respectively, while GPT-5 decreases by 11.1 percentage points. This difference may be related to the structure of the two benchmarks. The primary benchmark includes synthetic edge cases and hard negatives intended to stress safe/unsafe discrimination, whereas the external benchmark is based on more natural public datasets. Under these conditions, GPT-5 appears more prone to misclassifying ambiguous content when applying the shared taxonomy-constrained prompt. Beam q4_K_M shows much smaller variation across benchmarks, with a decrease of only 1.2 percentage points, suggesting stronger cross-benchmark stability.

The relative stability of model performance across the primary and external benchmarks is visualized in Figure 5.

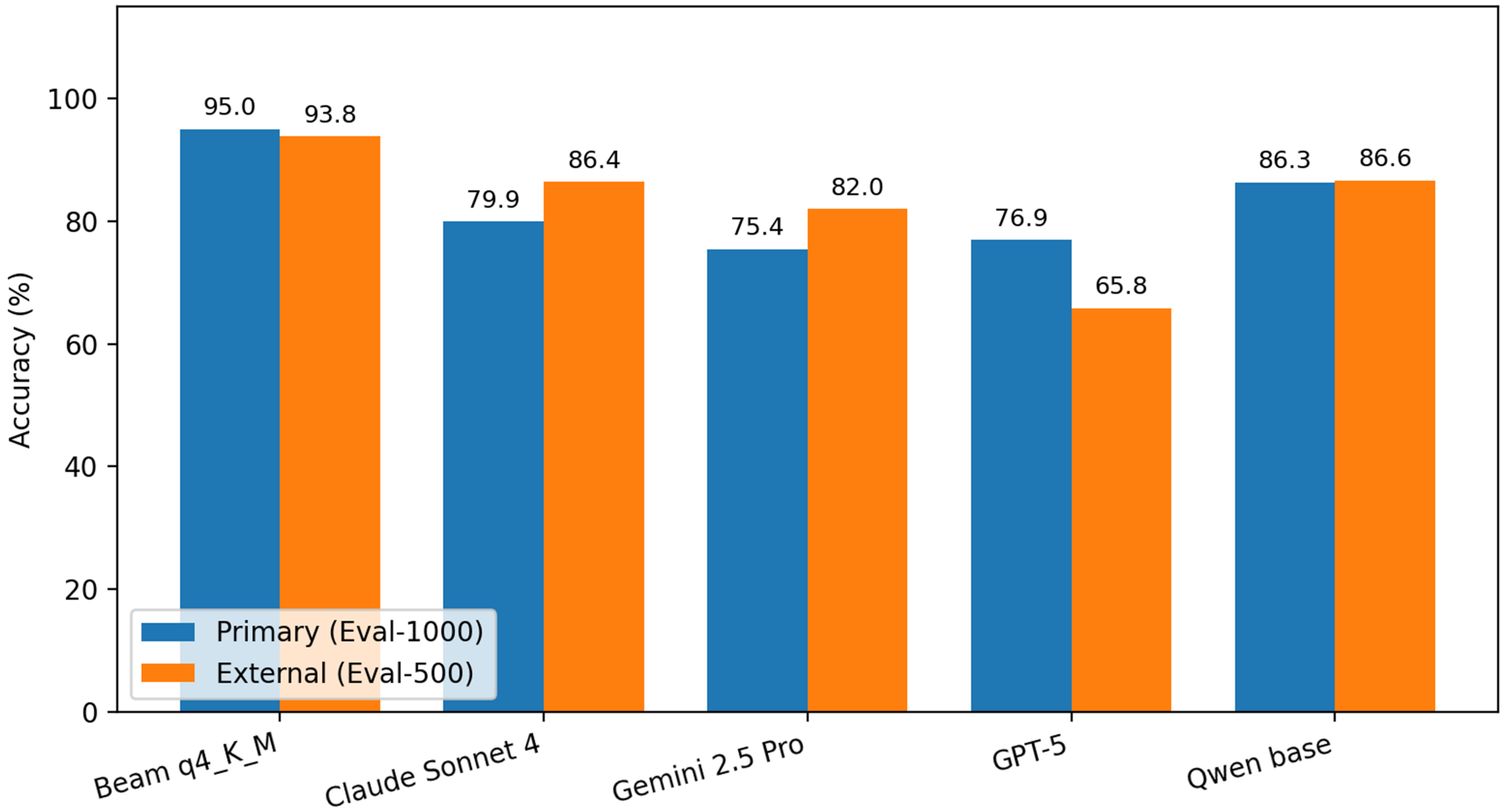

**Figure 5.** Accuracy of selected models on the primary (Eval-1000) and external (Eval-500) benchmarks.

**Source:** created by the author based on comparative results from the primary and external benchmarks.

Taken together, the primary and external benchmark results suggest that Beam q4_K_M's advantage is not confined to the synthetic evaluation setting. Its performance remains higher than the tested commercial baselines on held-out external data, supporting the claim of cross-dataset robustness. However, the margin over Claude Sonnet 4 narrows to approximately 7 percentage points on the external benchmark, indicating that the advantage is less pronounced on more naturalistic public datasets than on the primary stratified benchmark.

### 7.6. Manual reviewer validation

To check the reliability of the benchmark labels, we added a manual validation step. Two independent reviewers re-annotated a stratified subset of 200 evaluation samples, including 100 samples from the primary benchmark and 100 samples from the external benchmark (seed = 2026). The reviewers worked without access to the original benchmark labels or model outputs. Agreement at the combined category-and-subcategory level was high: 197 of 200 cases matched, giving 98.5% agreement and Cohen's $\kappa = 0.984$. We then conducted a second review of 18 boundary cases where the model prediction and benchmark label differed in a way that could indicate a benchmark-label error. In the full 200-sample subset, there were 25 model–benchmark disagreements. Of these, both reviewers confirmed 15 cases (60%) as benchmark labelling errors, while 10 cases (40%) were retained as model errors. After adjudication, Beam q4_K_M achieved 90.8% accuracy and 89.8% macro-F1 on the retained sample set ($n = 195$, after excluding five ambiguous cases). The blind review worksheets and adjudication keys are released with the benchmark dataset on Hugging Face.

## 8. Discussion and Comparison with Related Research

### 8.1. Comparison with prior work

Two prior studies provide the closest comparators. Huang (2025) showed that domain-specific fine-tuning of foundation LLMs substantially improves performance on cybersecurity tasks relative to prompted general-purpose models, with LoRA and QLoRA reaching results close to full supervised fine-tuning. Our +8.7 pp category-level gain (Beam q4_K_M vs the properly-prompted Qwen 3.5 27B base) and +29.2 pp gain on subcategory adherence are consistent with that pattern, and extend it to a 51-subcategory taxonomy in which schema adherence — rather than category discrimination — is the dominant contribution of fine-tuning.

### 8.2. Where the present results diverge

Zacharis et al. (2025) reported strong results for prompted frontier LLMs (GPT-4o, Gemini) on cybersecurity threat-extraction tasks. By contrast, the present evaluation shows frontier models trailing the fine-tuned local model by 15–20 pp on the fixed-taxonomy primary benchmark and by 7.4 pp on real-data external validation. The discrepancy is reconciled by task structure: threat extraction tolerates loosely-formatted output, whereas the present benchmark requires strict adherence to a 51-subcategory schema and a fixed JSON output protocol. Under those constraints, prompt-only general-purpose models — including GPT-5, despite its reasoning emphasis — produce schema-violating outputs (e.g., GPT-5 invents subcategory names such as credentials.aws_access_key_id), and category calibration becomes more important than extended reasoning capacity.

### 8.3. Threats to validity and limitations

We note five main threats to validity and limitations. First, Beam was evaluated with the same system prompt used during training, whereas the commercial models received this prompt only at test time. This may have placed the baseline models at a disadvantage. At the same time, we consider this setup realistic, since production systems typically rely on a fixed prompt. Even so, we acknowledge that model-specific prompt optimization for the baselines, including few-shot prompting or chain-of-thought prompting, might have reduced the observed gap. For this reason, the comparison should be understood as an evaluation of out-of-the-box adherence to a shared taxonomy and output protocol, rather than as a comparison based on vendor-optimized prompting. Second, the primary benchmark consists of synthetic documents created by the same author who developed the training corpus. Although the benchmark uses sources disjoint from training or explicitly held-out splits, distributional overlap between synthetic training and synthetic evaluation data cannot be fully excluded. This concern is only partially mitigated by the held-out external validation reported in Section 7.5, where Beam achieved 93.8% accuracy on 500 samples, including MTSamples (82.0%), a dataset explicitly excluded from training. Third, GPT-5 does not support deterministic inference (temperature = 0 is not available), so its accuracy may vary across runs. Ideally, multiple runs with confidence intervals would strengthen the comparison; we report a single run due to cost constraints ($10 per 1,000 samples).
Fourth, although category-level accuracy is strong, subcategory accuracy is lower (48.2% for Beam q4_K_M), indicating that fine-grained distinctions within the 51-subcategory taxonomy remain more challenging than top-level classification — a direction for future refinement. Fifth, the present study focuses on English-language documents and a single-label evaluation protocol; multilingual evaluation and multi-label scoring are left as future work.

## 9. Conclusion

TorchSight is an open-source on-premise security document classifier built around a Qwen 3.5 27B model fine-tuned with LoRA on a 78,358-sample multi-source corpus. Evaluation uses two complementary benchmarks: TorchSight-Eval-1000 (synthetic, stratified across all seven taxonomy categories) and TorchSight-Eval-500-External (real public datasets covering four categories where public data exists).

On the primary benchmark, Beam q4_K_M reaches 95.0% category-level accuracy and 48.2% subcategory accuracy, outperforming Claude Sonnet 4 (79.9%), Claude Opus 4 (79.9%), GPT-5 (76.9%), and Gemini 2.5 Pro (75.4%) by 15–20 percentage points and improving over the properly-prompted base Qwen 3.5 27B by 8.7 pp on category accuracy and 29.2 pp on subcategory adherence. The advantage holds on real public data: 93.8% on Eval-500-External, 7.4 pp ahead of the strongest commercial baseline. On safe documents, Beam's false-positive rate is 2.0% versus 22–63% for commercial APIs, suggesting that the model is operationally viable for DLP screening where commercial LLMs are not.

These results should be read in light of the limitations in Section 8.3 — synthetic training data for three categories with no public benchmark, single-prompt comparison, and a single-run protocol for GPT-5 (which does not support deterministic inference). To support independent verification, the full system is publicly released under Apache 2.0: benchmark data and per-model evaluation outputs at huggingface.co/datasets/torchsight/cybersecurity-classification-benchmark, Beam weights at huggingface.co/torchsight/beam-{q4_K_M,q8_0,f16}, and source code at github.com/IvanDobrovolsky/torchsight.

**10. Data Availability**

To ensure transparency, reproducibility, and practical adoption, all core components of TorchSight are publicly released under the Apache 2.0 license. This includes the training dataset comprising 78,358 curated samples (TorchSight, n.d.-a), the evaluation benchmark TorchSight-Eval-1000 with ground truth (TorchSight, n.d.-b), the Beam model weights in three quantizations – q4_K_M (TorchSight, n.d.-c), q8_0 (TorchSight, n.d.-d), and f16 (TorchSight, n.d.-e) – as well as the full system source code for both the CLI and desktop application (TorchSight, n.d.-f). These materials are publicly available and enable independent verification, reproducible benchmarking, and real-world deployment of the proposed approach.

**11. Funding**

This research did not receive any specific grant from funding agencies in the public, commercial, or not-for-profit sectors.

**12. CRediT authorship contribution statement**

Ivan Dobrovolskyi: Conceptualization, Methodology, Validation, Formal analysis, Investigation, Data curation, Writing – original draft, Writing – review & editing, Visualization, Project administration.

**13. Declaration of generative AI and AI-assisted technologies in the writing process**

During the preparation of this work, GPT-4 (OpenAI) was used as part of the research process to generate synthetic and hard-negative training samples, as described in Section 4.3. Claude Code Opus 4.6 (Anthropic) was used only for code assistance during software implementation and debugging. The author reviewed, tested, and verified the generated samples, code outputs, analyses, and manuscript content as needed and takes full responsibility for the content of the publication. No generative AI or AI-assisted tool was used to formulate the research idea, interpret the results, draw the conclusions, write the manuscript, or create or modify the figures.

**14. Declaration of competing interest**

The author is the developer and maintainer of TorchSight, the open-source system evaluated in this study. The author declares no known financial competing interests or personal relationships that could have appeared to influence the work reported in this paper.